# Evaluation of latent variances in Monte Carlo dose calculations with Varian TrueBeam photon phase-spaces used as a particle source

Eyad Alhakeem,[1, 2], Sergei Zavgorodni [2,1]

[1]Department of Physics and Astronomy, University of Victoria, Victoria, British Columbia V8W 3P6, Canada
[2]Dpartment of Medical Physics, British Columbia Cancer Agency–Vancouver Island Centre, Victoria, British Columbia V8R 6V5, Canada

Email: eyadali@uvic.ca

**Purpose:** To evaluate the latent variance (LV) of Varian TrueBeam photon phase-space files (PSF) for open 10x10 cm$^2$ and small stereotactic fields and estimate the number of phase spaces required to be summed up in order to maintain sub-percent latent variance in Monte Carlo (MC) dose calculations.

**Method:** BEAMnrc/DOSXYZnrc software was used to transport particles from Varian phase-space files (PSF$_A$) through the secondary collimators. Transported particles were scored into another phase-space located under the jaws (PSF$_B$), or transported further through the cone collimators and scored straight below, forming PSF$_C$. Phase-space files (PFS$_B$) were scored for 6MV-FFF, 6MV, 10MV-FFF, 10MV and 15MV beams with 10x10 cm$^2$ field size, and PSF$_C$ were scored for 6MV beam under circular cones of 0.13, 0.25, 0.35, 1.0, 1.2, 1.5 and 4 cm diameter. Both PSF$_B$ and PSF$_C$ were transported into a water phantom with particle recycling number ranging from 10 to 1000. For 10x10 cm$^2$ fields 0.5x0.5x0.5cm$^3$ voxels were used to score the dose, whereas the dose was scored in 0.1x0.1x0.5cm$^3$ voxels for beams collimated with small cones. In addition, for small 0.25 cm diameter cone-collimated 6MV beam, phantom voxel size varied as 0.02x0.02x0.5cm$^3$, 0.05x0.05x0.5cm$^3$ and 0.1x0.1x0.5cm$^3$. Dose variances were scored in all cases and latent variance evaluated as per Sempau *et al*.

**Results:** For the 10x10 cm$^2$ fields calculated LVs were greatest at the phantom surface and decreased with depth until reached a plateau at 5 cm depth. LVs were found to be 0.54%, 0.96%, 0.35%, 0.69% and 0.57% for the 6MV-FFF, 6MV, 10MV-FFF, 10MV and 15MV energies, respectively at the depth of 10cm. For the 6 MV phase-space collimated with cones of 0.13, 0.25, 0.35, 1.0 cm diameter, the LVs calculated at 1.5 cm depth were 75.6%, 25.4%, 17.6% and 8.0% respectively. Calculated latent variances for the 0.25 cm cone-collimated 6MV beam were 61.2%, 40.7%, 22.5% in 0.02x0.02x0.5cm$^3$, 0.05x0.05x0.5cm$^3$ and 0.1x0.1x0.5cm$^3$ voxels respectively.

**Conclusions:** In order to achieve sub-percent latent variance in open 10x10 cm$^2$ field MC simulations single PSF can be used, whereas for small SRS fields more PSFs would have to be summed.

1.  **INTRODUCTION**

Monte Carlo (MC) simulation of a radiotherapy beam is often carried out as a two- stage process. The first stage involves modeling the invariable, plan-independent upper part of a radiotherapy linear accelerator (linac) head. Particles simulated at this stage are then scored into a phase-space located at a pre-defined surface in the head model. The PSF contains particle fluence information (coordinates, directional cosines, energy, type of particle) of the modeled beam. At the second stage, the PSF is used as a particle source and MC transport code(s) propagate the particles through plan-dependent part of the linac head into a phantom.

In order for a phase-space file to allow accurate dose calculations in the phantom, it has to reflect particle fluence of the real radiation beam. This can be achieved by thoroughly modeling parameters of the initial electron beam that hits the target as well as beam-shaping components of the linac head. In the past, Varian (Varian Medical Systems, Palo Alto, CA) allowed access to schematics of their linacs to facilitate modeling of particle transport for the dose calculations. This provided gateway for modeling linac head and generating phase spaces with as many particles as required. For the TrueBeam linac, Varian did not release the treatment head schematics; instead they provided PSFs, containing about $50 \times 10^6$ particles each, to be used in Monte Carlo calculations. TrueBeam version 1 phase-space files were scored on a curved surface and contained 2-4 GB data per photon energy. Varian later released version 2 of the TrueBeam phase-space files. Version 2 phase-spaces were further tuned to obtain better agreement with representative set of "golden beam data" measurements, and they were scored on flat surface. Version 2 library contains fifty files of ~1 GB size per photon energy.

It has been long established that variance in the MC calculated dose is inversely proportional to the particle density in the fluence incident on a phantom (Mackie, 1996). Later, Sempau *et al* (2001) showed that regardless of variance reduction techniques used in the PSF particles' transport through a phantom the variance of a calculated quantity cannot be reduced below a certain limit. This smallest possible value of the variance was named the latent variance. Sempau *et al* (2001) also proposed a technique for determining the value of latent variance.

Latent variance therefore limits the accuracy achievable in MC dose calculations that use PSFs as a particle source, and in order to reduce latent variance larger phase-space files have to be used. Still, very large PSFs are awkward to use as they take large amounts of storage space and prone to errors when transported through networks. On the other hand, if the latent variance of these phase-spaces has not been quantified, there is no guide on how many of them need to be used to achieve required accuracy. In fact, some applications are likely to require summing up more phase-space files than the others. This would depend on the energy of the beam, field size, and the required dose grid resolution. Cronholm and Behrens (2013) in a conference abstract reported evaluation of latent uncertainties for version 1 Varian TrueBeam 6MV-FFF, 6MV, 10MV-FFF and 10MV phase-space files and $10 \times 10$ cm$^2$ fields. However, this abstract did not provide details of their study nor explored the effect of depth, field size and voxel size on latent uncertainties in the calculated dose.

This study provides an evaluation of the latent variances for version 2 of Varian TrueBeam photon PSFs. Latent variances were evaluated at different depths in a phantom for various beam energies and phantom grid resolution. Standard $10 \times 10$ cm$^2$ fields as well as small and very small fields (down to 0.13 cm diameter) were included. In addition, estimation on the number of 6 MV phase-space files, or particles, required to achieve sub-percent latent uncertainty is provides for various field sizes.

2.  **METHODS AND MATERIALS**

*2.1 Phase-space files for the Varian TrueBeam linacs*

Varian (Varian Medical Systems, Palo Alto, California) research team generated TrueBeam PSFs by modeling particle transport through the linac treatment head (Constantin *et al* 2011) with Monte Carlo GEANT4 (Agostinelli *et al* 2003) code. They used computer aided design schematics of the linac head and integrated these schematics into GEANT4 geometry modules. Simulated particles were scored between the shielding collimator and the upper jaws in an International Atomic Energy Agency (IAEA) phase-space file format (Capote *et al* 2006). Fifty PSFs per beam energy were generated for version 2 phase-space library and each PSF contains about 50 million particles producing a binary file of about 1 GB size. This required the numbers of electrons incident on the target ($N_i$) to be in the range of $3 \times 10^8 - 9 \times 10^8$ (as shown in table 1) depending on the beam energy.

*2.2 Evaluation of latent variance*

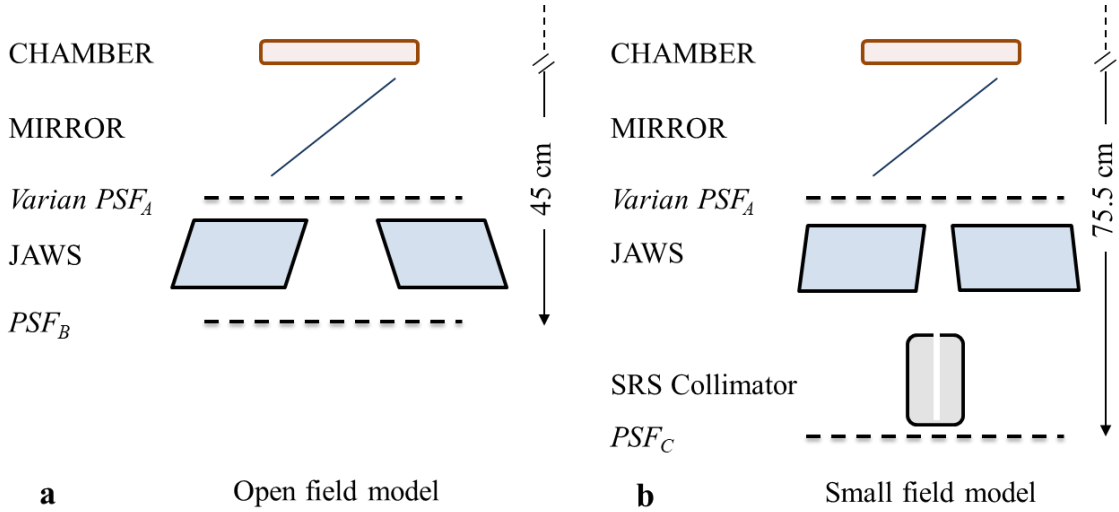

**Figure 1.** Diagram illustrating BEAMnrc models used in this study. Shown are Varian PSF (PSF$_A$), and a) the PSF$_B$ scored under secondary collimator; b) the PSF$_C$ scored under the SRS collimator.

We modelled open 10x10 cm² fields and fields collimated by a circular stereotactic collimator as shown in figure 1. For these fields latent variances of Varian TrueBeam PSFs were evaluated using technique similar to that proposed by Sempau *et al* (2001).

Sempau *et al* (2001) evaluated latent uncertainty of a small PSF by using the following technique. They run a few simulations transporting particles from the PSF into a phantom with various particle splitting factors K and then plotted the variance of the scored quantity against $K^{-1}$. From the linear relation between the two, latent variance $\sigma_K$ was obtained as the linear-fit intercept at $K^{-1}=0$. To obtain the latent variance of a typical PSF, $\sigma_K$ was scaled down by the ratio of particles in the two PSFs.

In our study one of Varian PSFs (called PSF$_A$ further in the text) was transported through the collimators and additional PSFs (called PSF$_B$ and PSF$_C$) were scored as shown in figure 1. These ancillary phase-spaces contained a smaller number of particles and required shorter simulation time. We therefore initially evaluated latent variances of PSF$_B$ and PSF$_C$ using technique by Sempau *et al* (2001) and then scaled these variances to obtain the latent variance of PSF$_A$ as detailed below. The number of particles in PSF$_B$ and PSF$_C$ varied depending on the field size and are shown in table 1. Procedures for evaluating PSF$_B$ and PSF$_C$ are identical, therefore only that for PSF$_B$ is described.

**Table 1**. A summary of the information for Varian PSFs investigated in this work. Where, $N_i^A$ is the number of electrons incident on the target to create PSF$_A$, $N_{PSA}$ is total number of particles in one full

PSF$_A$, $N_i^B$ (or $N_i^C$) is the number of electrons incident on the target to create PSF$_B$ (or PSFc); and $N_{PSB}$ (or $N_{PSC}$) is the total number of particles in PSF$_B$ (or PSFc).

|  | Beam energy/Field size | $N_i^A$ | $N_{PSF_A}$ | $N_i^B, N_i^C$ | $N_{PSF_B}, N_{PSF_C}$ |
|---|---|---|---|---|---|
| | 6 MV/10x10 cm² | 9.0 ×10⁸ | ~ 46×10⁶ | 9.7×10⁷ | 255,298 |
| **Open fields** | 6 MV-FFF/10x10 cm² | 6.5×10⁸ | ~ 47×10⁶ | 3.2×10⁷ | 202,433 |
| | 10 MV/10x10 cm² | 5.2×10⁸ | ~ 49×10⁶ | 4.69×10⁷ | 210,159 |
| | 10 MV-FFF/10x10 cm² | 3.24×10⁸ | ~ 45×10⁶ | 1.62×10⁷ | 240,340 |
| | 15 MV/10x10 cm² | 6.0×10⁸ | ~ 48×10⁶ | 5.53×10⁷ | 224,274 |
| **Small SRS fields** | 6 MV/Cone 1.0cm | 9.0×10⁸ | ~ 46×10⁶ | 9.0×10⁸ | 17,639 |
| | 6 MV/Cone 0.35cm | 9.0×10⁸ | ~ 46×10⁶ | 9.0×10⁸ | 5,338 |
| | 6 MV/Cone 0.25cm | 9.0×10⁸ | ~ 46×10⁶ | 9.0×10⁸ | 6,751 |
| | 6 MV/Cone 0.13cm | 9.0×10⁸ | ~ 46×10⁶ | 9.0×10⁸ | 3,780 |

DOSXYZnrc code (Walters *et al* 2005) was used to transport PSF$_B$ into a phantom and score the dose and its variance. MC simulations were then repeated with an increased number of particle recycling ($N_{recycl}$). The value of dose variance in a voxel located at the beam Central Axis (CAX) was calculated from each simulation, and latent variance (LV$_B$) was obtained by extrapolating the variances to infinite recycling number ($1/N_{recycl}$ equal to zero) as seen in LV evaluation plots (figure. 2).

The latent variance of a phase-space is inversely proportional to the number of particles it contains as $LV \propto 1/N_{PSF}$. Therefore, in order to evaluate the latent variance LV$_A$ of a full PSF$_A$, the latent variance values, LV$_B$ were scaled to the number of incident electrons (shown in table 1), that have been used to generate a single PSF$_A$ as:

$$LV_A = \frac{N_i^B}{N_i^A} \times LV_B \quad (1)$$

Where $N_i^A$: the number of electrons incident on the target to create a full Varain PSF$_A$
$N_i^B$ : the number of electrons incident on the target to create PSF$_B$
$LV_B$: calculated latent variance of PSF$_B$.

Using equation (1) and the calculated latent variances of Varian PSFs ($LV_A$) for different beam energies, expressed in percent, total number of particles in a phase-space to achieve latent variance of 1% ($N_{PSF}^{1\%}$) can be calculated as:

$$N_{PSF}^{1\%} = N_{PSF_A} \times \frac{LV_A}{1\%} \quad (2)$$

Then the number of phase-spaces M required to be summed to achieve a 1% latent variance can be written as $M = \frac{N_{PSF}^{1\%}}{N_{PSF_A}}$. , and from equation 2:

$$M = LV_A(\%) \quad (3)$$

Therefore the number of phase spaces that need to be summed to produce a single phase space with

$LV=1\%$ is simply equal to the value of $LV_A$, expressed in percent.

In all considered cases, latent variances were evaluated in voxels at the CAX of the beam. For 10x10 cm² 6MV-FFF, 6MV, 10MV-FFF, 10MV and 15MV beams LVs were evaluated at depths ranging from 0.25 cm to 15 cm in water phantom with 0.5x0.5x0.5 cm³ voxels. For small 0.13m, 0.25cm, 0.35cm, and 1.0cm diameter 6MV beams latent variances were evaluated in 0.1x0.1x0.5 cm³ voxels. In addition, latent variances in voxels of 0.02x0.02x0.5 cm³, 0.05x0.05x0.5 cm³ and 0.1x0.1x0.5 cm³ were evaluated for the 0.25 cm diameter SRS cone. Latent variances in this case were determined at 1.5 cm depth for all voxel sizes, as for small SRS fields LV was found to be minimal at this depth rather than at the depth of 10 cm.

Evaluation of uncertainties in this work only includes statistical uncertainty (known as type-A uncertainty (Andreo and Fransson 1989)) in the calculated dose. Other possible uncertainties that arise from inaccuracies in linac model geometry, approximations build into MC particle interaction models, and cross sections (known as type B uncertainties (Andreo and Fransson 1989)) were not considered in this work.

## 3. RESULTS

The latent variance evaluation plots for different energies are shown in figure 2, and the latent variance values as determined at different depths in a water phantom are shown in table 2.

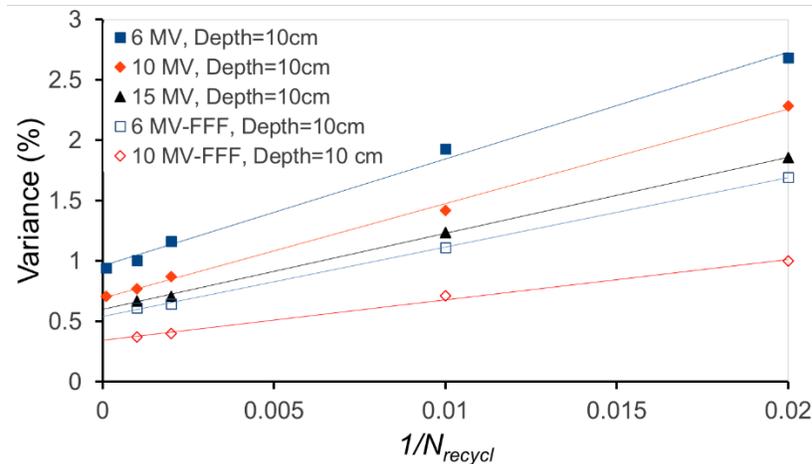

**Figure 2**. Latent variance evaluation plots (dose variance vs $1/N_{recycl}$) for 6MV-FFF, 6MV, 10MV-FFF, 10MV and 15MV 10x10 cm² open fields. Variances were scored at 10 cm depth in a water phantom with 0.5x0.5x0.5 cm³ voxels.

**Table 2**. Calculated latent variance values for 6MV-FFF, 6MV, 10MV-FFF, 10MV and 15MV 10x10 cm² open fields, in 0.5x0.5x0.5 cm³ voxels at different depths in water.

| | Latent variance (%) | | | | |
|---|---|---|---|---|---|
| | Energies (MV) | | | | |
| **Depth, cm** | 6 | 10 | 15 | 6 MV-FFF | 10 MV-FFF |
| **0.25** | 1.39 | 1.41 | 21.49 | 0.73 | 0.56 |
| **1.5** | 1.05 | 0.92 | 0.57 | 0.61 | 0.36 |
| **10** | 0.96 | 0.69 | 0.57 | 0.54 | 0.35 |

As seen from table 2, at 10 cm depth latent variances are minimal and decrease as the beam energy increase; the latent variances for FFF beams are nearly twice smaller than those for flattened beams.

Latent variances, calculated for 6MV 10x10 cm$^2$ open field at different phantom depths are shown in figure 3. Calculated latent variances were found to be the highest at the surface, and decreased with depth until plateauing at about 5 cm depth.

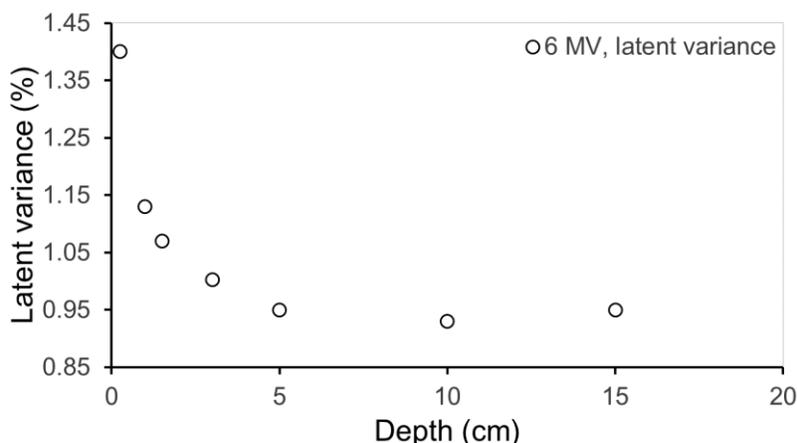

**Figure 3**. Latent variance, calculated for 6MV open 10x10 cm$^2$ field as a function of depth in a water phantom with 0.5x0.5x0.5 cm$^3$ voxels.

**Table 3.** Latent variance values for 6 MV SRS small fields evaluated at different depths in the phantom, in 0.1x0.1x0.5cm$^3$ voxels.

| | **Latent variance %** | | | |
|---|---|---|---|---|
| | Small SRS fields | | | |
| **Depth, cm** | Cone 1.3 | Cone 2.5 | Cone 3.5 | Cone 10 |
| **0.25** | 268.2 | 28.3 | 22.1 | 16.7 |
| **1.5** | 66.0 | 22.5 | 15.8 | 8.2 |
| **10** | 75.6 | 25.4 | 17.6 | 8.0 |

Latent variance values for the small fields at 0.25 cm, 1.5 cm, and 10 cm depths are shown in table 3. The evaluated latent variances increased as the cone size decreased, but unlike open 10x10cm$^2$ field, latent variances were minimal at the depth of 1.5cm.

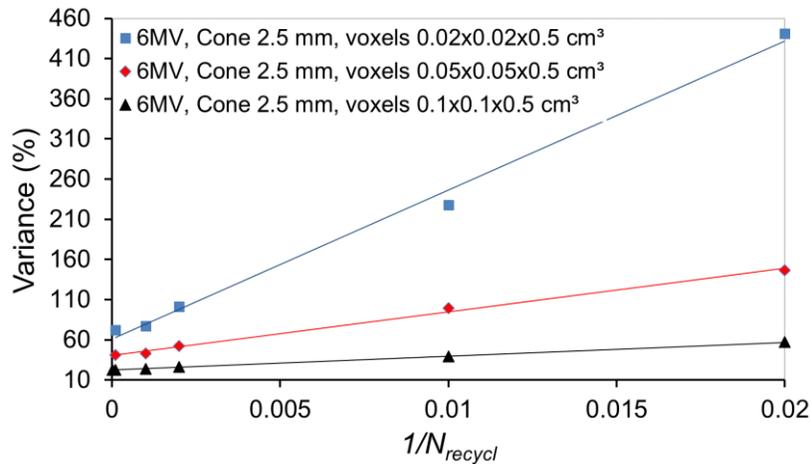

**Figure 4.** Latent variance evaluation plots (dose variance vs $1/N_{recyl}$) for 6 MV, 0.25cm field. Variances were scored in voxels of $0.02 \times 0.02 \times 0.5 cm^3$, $0.05 \times 0.05 \times 0.5 cm^3$ and $0.1 \times 0.1 \times 0.5 cm^3$ size, at 1.5 cm depth.

Latent variance evaluation plots for 0.25cm cone-collimated 6MV beams, with variance scored in different voxel sizes at 1.5 cm depth, are shown in figure 4. The latent variance values were found to be 61.2%, 40.7% and 22.5% for voxels of $0.02 \times 0.02 \times 0.5 cm^3$, $0.05 \times 0.05 \times 0.5 cm^3$ and $0.1 \times 0.1 \times 0.5 cm^3$ size, respectively.

Table 4 shows an estimated of number of Varian 6MV PSFs needed to achieve latent variances of 1% at 1.5 cm depth for the various cones investigated in this work, as well as for $10 \times 10 cm^2$ open field. As shown in the Methods the number of PSFs that needs to be summed up is equal to the value of latent variance expressed in percent.

**Table 4**. Estimated number of 6 MV Varian TrueBeam PSFs needed to achieve the latent variance of 1.0% at 1.5 cm depth. The variance was scored in $0.5 \times 0.5 \times 0.5 cm^3$ voxels for $10 \times 10 cm^2$ open field and in $0.1 \times 0.1 \times 0.5 cm^3$ voxels for the small cones.

|  | LV of a full TrueBeam PSF | Number of PSF's required to achieve 1.0% LV |
|---|---|---|
| $10 \times 10 cm^2$ (6MV) | 1.05% | ~ 1 |
| Cone 10 | 8.2% | ~ 8 |
| Cone 3.5 | 15.8% | ~ 16 |
| Cone 2.5 | 22.5% | ~ 23 |
| Cone 1.3 | 66.0% | ~ 66 |

## 4. DISCUSSION

To the best of our knowledge this is the first study evaluating the latent variance in the MC calculated dose for small and very small fields. The impact of field size and phantom voxel size on the latent variance was also evaluated. For standard $10 \times 10 cm^2$ fields latent variance was evaluated for different energies and depths in the phantom.

Our results for small fields demonstrate dramatic impact of latent variance on MC calculations for small field dosimetry. They show that even by summing up all fifty phase space files, currently available for TrueBeam linac, it would be impossible to reduce the latent variance below 1% for very small (less than 1.5mm) fields. For the calculations that require higher resolution the problem will be even more exacerbated and accurate calculations would only be achievable through using latent variance reduction techniques.

Latent variance reduction techniques that utilize circular symmetry of the phase space above secondary collimators have been previously proposed and investigated (Bush *et al* 2007, Fix *et al* 2004, Brualla and Sauerwein 2010). These techniques have shown the capability to reduce the latent variance by more than a factor of 20 (Bush *et al* 2007) for calculations with particle transport in Cartesian coordinate system. However, the reduction in PSF latent variance was shown to be inversely proportional to the radial distance (measured in voxel sizes) from the beam central axis. Thus, such variance reduction techniques would not be efficient for modelling applications that require high accuracy in a voxel at CAX, and summing up of many PSF's would be required.

Walters et al. (2002) investigated the effect of recycling on uncertainty in "particle-by-particle" method of uncertainty estimate. For an 18MeV electron beam ($20 \times 20 cm^2$, SSD=100cm) they found that when recycling number was largest (twenty seven), uncertainty was high at the surface and decreased with depth until reached the minimum at the depth of about 5 cm. Uncertainty increased gradually beyond that depth. The increase of uncertainty at surface was attributed to the fact that scored quantities were grouped by contributions from primary histories. A primary history was defined as MC trajectory (that includes all secondary generated particles) initiated from the initial electron entering the linac head and all occurrences of this particle created due to recycling. The number of contributions from primary histories that determines uncertainty in a voxel at surface was smaller than that at 5cm depth because at depth more primary particles reached the voxel and interacted due to their scatter from wider area, Similar to the results by Walters *et al.* (2002), latent variances in our study were largest at the surface, and this behavior, as well as the behavior of latent variance with field size, voxel size, beam energy can be explained through the number of primary particles contributing to the dose in a voxel. The number of contributions will indeed increase with the field or voxel size, and it will increase with beam energy due to more forward directed particle fluence. Likewise, un-flattened beams are also more forward directed producing more interaction in a voxel at the central axis and subsequently lower latent variance.

Our results show that for $10 \times 10$ cm$^2$ open fields simulations with different beam energies, sub-percent latent variance can be achieved with a single PSF as the evaluated latent variances range from 0.35% to 1%. This is consistent with the findings by Cronholm and Behrens (2013). They evaluated latent uncertainties of Varian TrueBeam version 1 phase-spaces for $10 \times 10$ cm$^2$ field and $0.25 \times 0.25 \times 0.25$ cm$^3$ voxels located near the beam isocenter, and found them to be 0.85%, 1.02%, 0.41% and 0.74% for the 6 MV-FFF, 6 MV 10 MV-FFF and 10 MV photon beams, respectively. This study was a conference abstract and did not contain substantial details of the calculations.

## 5. CONCLUSIONS

The number of PSFs that need to be summed up was evaluated and provided in this study and shown to be equal numerically about equal to the value of latent variance evaluated for the conditions of the simulation.

A single phase space should be sufficient to achieve sub-percent latent variance for 10x10 cm2 fields in 0.5x0.5x0.5cm3 voxels when using Varian TrueBeam PSFs. However many PSFs would have to be summed up for accurate small field MC calculations.